# Room temperature quantum emitters in van der Waals $\alpha$-$MoO_3$

*Jeonghan Lee,*[1,2#] *Haiyuan Wang,*[3#] *Keun-Yeol Park,*[1] *Soonsang Huh,*[1,2] *Donghan Kim,*[1,2] *Mihyang Yu,*[1,2] *Changyoung Kim,*[1,2] *Kristian Sommer Thygesen,*[3*] *and Jieun Lee,*[1,2*]

[1]Department of Physics and Astronomy, Seoul National University, 08826 Seoul, Korea

[2]Center for Correlated Electron Systems, Institute for Basic Science, 08826 Seoul, Korea

[3]Computational Atomic-Scale Materials Design (CAMD), Department of Physics, Technical University of Denmark, 2800 Lyngby, Denmark

[#] These authors contributed equally to this work.

[*] To whom correspondence should be addressed, lee.jieun@snu.ac.kr, thygesen@fysik.dtu.dk

## Abstract

**Quantum emitters in solid-state materials are highly promising building blocks for quantum information processing and communication science. Recently, single-photon emission from van der Waals materials has been reported in transition metal dichalcogenides and hexagonal boron nitride, exhibiting the potential to realize photonic quantum technologies**

**in two-dimensional materials. Here, we report the generation of room temperature single-photon emission from exfoliated and thermally annealed single crystals of van der Waals *α*-$MoO_3$. The second-order correlation function measurement displays a clear photon antibunching, while the luminescence intensity exceeds 0.4 Mcts/s and remains stable under laser excitation. The theoretical calculation suggests that an oxygen vacancy defect is a possible candidate for the observed emitters. Together with photostability and brightness, quantum emitters in *α*-$MoO_3$ provide a new avenue to realize photon-based quantum information science in van der Waals materials.**

Quantum emitters are central resources for a variety of quantum technologies [1], including quantum sensing [2], computing [3], and communication [4]. Single-photon emitters have been reported in solid-state systems like color centers in diamond [5], silicon carbide [6], quantum dots [7], and recently in van der Waals materials [8-12]. Due to the proximity to the surrounding environment, 2D quantum emitters enable precise and robust light-matter interactions at the nanoscale. Furthermore, combined with the capability to be integrated into heterostructures by stacking various materials, van der Waals quantum emitters have established a promising platform for optoelectronics and chip-based quantum technologies. For instance, quantum light sources in TMDCs can be readily activated through strain [13] and moiré confinement [14], while inheriting unique chiral optical properties from the host materiel's valley and spin degrees of freedom [15]. *h*-BN, a wide bandgap van der Waals material, has also been extensively studied for the room temperature generation of bright single-photon emitters [16, 17]. Recent advancements have also highlighted the exceptional potential of 2D spin defects, including the realization of quantum sensing at atomic proximity through optically addressable localized spins in *h*-BN [18, 19]. However, quantum emitters in van der Waals materials have been reported in only a few material platforms so far [20], urging the necessity to explore single-photon emitters in a wider material range. Expanding the diversity of material platforms of 2D quantum emitters will propel the field forward and unlock novel applications in quantum photonics landscape.

$\alpha$-$MoO_3$ is a van der Waals semiconductor with a bandgap of 3 eV [21], which makes it a promising candidate for generating room temperature single-photon sources. Moreover, the defects in $\alpha$-$MoO_3$ have been shown to give rise to intriguing physical phenomena. Notably, studies have demonstrated that controlling oxygen vacancies can induce intermediate gap states, enhancing carrier mobility and photo-responsivity significantly [22, 23]. Additionally, emissions associated with these defects have been observed in the wavelength range of 500 – 600 nm [24, 25], although only ensemble measurements have been reported so far. Recent studies also have suggested that unpaired electrons of oxygen vacancies in $\alpha$-$MoO_3$ can induce ferromagnetic alignment of defect spins even at room temperature [26, 27]. Thanks to the refractive index of approximately 2.1 [28] similar to that of silicon nitride [29], it can be envisaged that fully monolithic quantum devices conjoined with other quantum emitters in 2D platforms could be developed. Although single-photon emitters in *h*-BN and chalcogenide based van der Waals materials have been widely studied for their quantum properties, unveiling quantum emitters in $\alpha$-$MoO_3$ with a distinctive bandgap and electrical properties are likely to have a significant impact on 2D material based quantum photonics and optoelectronics.

In this work, we investigate the quantum optical properties of the localized emission from layered $\alpha$-$MoO_3$ and study the electronic and optical characteristics of single defects. The second-order correlation function measurement performed on localized emission confirms quantum light generation at room temperature. The zero-phonon lines of these emitters are found to be predominantly distributed in the energy range between 2.0 and 2.3 eV with high emission intensity. Also, surface treatments and mechanical stacking of van der Waals heterostructure formation are found to enhance the photon emission stability. By calculating the photoluminescence spectrum, we identify an oxygen vacancy as a highly probable origin of the observed quantum emitters. Our work thus demonstrates $\alpha$-$MoO_3$ as a novel platform to study isolated defects and single-photon emitters in layered materials for room temperature quantum technologies.

The crystal structure of $\alpha$-$MoO_3$ consisting of bilayers of covalently bonded $MoO_6$ octahedra is shown in Figure 1a. In the bulk form, $MoO_6$ octahedra bilayers are held together by weak van der Waals forces, forming an orthorhombic structure. The crystal in our work is grown by the Bridgeman method (details in Supporting Information 1) and characterized by high-resolution X-ray diffraction (XRD) measurement. Figure 1b shows the XRD data of an as-grown crystal with three peaks at 12.7°, 25.6°, and 38.9°, corresponding to the lattice plane indices of (020), (040), and (060), respectively [30]. In the XRD data, no other peaks were observed supporting the high crystallinity of the material.

We obtained thin $\alpha$-$MoO_3$ flakes by the mechanical exfoliation of a bulk crystal. Figure 1c shows the optical microscope image of an exfoliated $\alpha$-$MoO_3$ flake and the inset shows the thickness of the flake measured with atomic force microscopy (AFM). We note that the thickness of $\alpha$-$MoO_3$ flakes used in our experiment ranges between 3.5 and 30 nm. The thinnest layer sample we could obtain was 5-layer thick which showed the signature of a quantum emitter as well (Figure S1, Supporting Information). The flakes were then annealed in high temperature and low-pressure environment to produce optically active defect sites. We also note that it was possible to observe quantum emitters in some flakes without annealing (Figure S2), but the occurrence of the emitters was lower. Raman spectrum measurement performed before and after the annealing confirmed that the overall crystal structure remained intact under the annealing process (Figure S3).

For optical measurements, we used a home-built scanning confocal microscopy set-up for photoluminescence and photon correlation experiments (Supporting Information 1). By using a sub-bandgap continuous-wave laser centered at 532 nm (2.3 eV), spatially localized emission is measured from several different spots on $\alpha$-$MoO_3$ flakes. A representative emission at room temperature is shown in Figure 1d with the 2D spatial map of confocal PL scan around the emission. We confirmed the single-photon characteristics of the observed emitter by the second-order correlation function $g^2(\tau)$ detection using the Hanbury-Brown-Twiss (HBT) interferometer.

To extract information from the $g^2(\tau)$ curve, we fitted the correlation function using a three-level model formula,

$$g^2(\tau) = 1 - p\left[(1+a)e^{-|\tau|/\tau_1} - ae^{-|\tau|/\tau_2}\right], \quad (1)$$

where $a$ and $p$ are fitting parameters while $\tau_1$ and $\tau_2$ are the lifetime of the excited and meta-stable states, respectively. The extracted excited state lifetime was 3.95 ns and meta-stable state lifetime was 0.76 μs, representing that the transition rate of non-radiative decay channel is much lower than the radiative decay rate. After the background correction following the procedure shown in Figure S4 [31], the $g^2(0)$ of this emitter was 0.16 ± 0.07, demonstrating the room temperature quantum emitting properties from local defects of $\alpha$-$MoO_3$. Figure S5 shows that the emitters also exhibit stable emission properties at room temperature.

We then measured various optical properties of quantum emitters in $\alpha$-$MoO_3$ at the temperature of 10 K. Figure 2a shows the spectrum of multiple emitters at low temperature, showing detailed spectral features such as zero-phonon line (ZPL) and multiple phonon side bands (PSB) at longer wavelengths. The ZPL energies of these emitters were found at 563, 572, 604, and 626 nm, respectively. It is also observed that the PSB of all emitters are separated by ~160 meV from the ZPL. The antibunching characteristics are verified for all emitters (Figure S6) as exemplified by the emitter B shown in Figure 2b. The lowest $g^2(0)$ value measured at cryogenic temperature was 0.087 which is shown in Figure S4. Surprisingly, we also observed an emitter with resolution-limited linewidth as narrow as 72 GHz (Figure S7). Such narrow linewidth implies the potential for coherent light generation and intrigues further study such as resonant excitation of an emitter [32] to acquire comprehensive view of the dephasing mechanisms for investigating quantum interference [33].

Another interesting point is that most of the emitters that we observed are found to have ZPL energy predominantly around 2.2 eV (Figure 2c). From the Gaussian fitting of the distribution of the emission energy, the central ZPL energy was 2.18 eV with standard deviation of 230 meV. The inhomogeneous ZPL distribution may have several origins such as strain or edge effect around the defect site or charged impurities in the nearby environments [34]. However, a rather narrow range of ZPL distribution suggests the high possibility to identify the defect structure and achieve homogeneous emitter generation. Given that e-beam irradiation has shown a narrow distribution of emitter energies down to a few nm in $h$-BN [32, 35], we anticipate that optimizing defect generation could also improve the ZPL energy homogeneity in $\alpha$-$MoO_3$.

Furthermore, by calculating the Debye-Waller factor of measured emitters, we found that the average ratio of the ZPL emission to total emission was about 0.2, showing that a substantial portion of the emission is channeled into the ZPL. From these emitters, we estimated the average Huang-Rhys (HR) factor to be about 1.9 (Figure S8). In addition, we measured $g^2(\tau)$ at various excitation powers of an emitter to investigate the photophysics and dynamics of the emission to obtain emitter quantum efficiency (Figure S9).

By performing the power dependence measurement, we also determined the saturation intensity of the luminescence of an emitter. In Figure 2d, the emission count rate ($I$) is depicted as a function of the laser excitation power ($P$). The experimental data was fitted using the following equation,

$$I(P) = I_{\mathrm{sat}} \frac{P}{P + P_{\mathrm{sat}}} \tag{2}$$

where $I_{\mathrm{sat}}$ and $P_{\mathrm{sat}}$ are the maximal emission count rate and the saturation power, respectively. The fitting result shows that $P_{\mathrm{sat}}$ and $I_{\mathrm{sat}}$ are 199 μW and 0.41 Mcts/s, respectively. A table summarizing emission intensity including quantum emitters in other van der Waals materials is shown in Supporting Information 11. Considering the collection efficiency (~ 0.027) of our set-up along the path to the spectrometer, the actual $I_{\mathrm{sat}}$ is calculated to be about 15 Mcts/s. Moreover, the integrated emission intensity was not found to decrease as temperature increases as presented in Figure S10. This feature is speculated to originate from complex excited-state orbital interactions, similar to that observed for NV centers [36]. We expect that through integration with optical cavities or plasmonic nanostructures, further improvements for light extraction can be made [37].

We then carried out several measurements to characterize the optical properties of $\alpha$-$MoO_3$ quantum emitters. As displayed in Figure 3a, we obtained the lifetime of an emitter using a time-resolved measurement with a pulsed laser excitation centered at 532 nm. The lifetime of the emitter was found to be about 3 ns, quantitatively agreeing with $\tau_1$ obtained from Figure 2b and Figure S6.

Also, the emitters from $\alpha$-$MoO_3$ flakes showed enhanced photostability after surface treatments using polymer coating [38] or integration with other van der Waals materials [39]. The spectrum and photostability of representative emitters from bare and polymer coated $\alpha$-$MoO_3$ flakes are shown in Supporting Information 13. Without surface treatment, in addition to emitters showing stable emission properties, there were some emitters exhibiting intermittent blinking of emission intensity. By using polymer coating treatment, significant enhancement of the emission stability could be achieved for the majority of emitters. Figure 3b and 3c show the fluorescence measurement result of an emitter processed with surface polymer coating showing the absence of photon blinking or bleaching. Our finding shows that instability in some emitters could have originated from photo-induced charge traps near the surface which could be suppressed by edge modulation. Other than polymer coating, integration of a few-layer graphene on top of an optically active defect site in $\alpha$-$MoO_3$ also effectively showed the enhancement of the emission stability. This further suggests the potential of charge state modulation of the observed emitters through van der Waals integration.

We also note that the atomic defects generated near the surface of exfoliated $\alpha$-$MoO_3$ confirm that the measured emitters are not from artifacts such as organic molecules at the interface between the substrate and host materials [40]. In Figure S14, we present detailed examinations including suspended sample fabrications, ruling out other possibilities as potential emitters.

Figure 3d and 3e shows the polarization dependence measurement of an emitter from which information on structural properties of the defect site can be extracted. The two-fold symmetry of the polarization dependence suggests the existence of structural anisotropy in the crystal structure of the defect. Closely aligned excitation and emission polarization orientations of the emitter (Figure 3f) further indicate that carriers are excited by phonon-mediated direct excitations.

To gain insight into the origin of the observed quantum emitters, we now turn to the theoretical investigation of the defect structure. As oxygen vacancies are known to induce intermediate-gap states in $\alpha$-$MoO_3$, it is highly probable that oxygen vacancy sites are responsible for the observed emitters. In the crystal structure of $\alpha$-$MoO_3$, there are three inequivalent oxygen sites: $O_1$, $O_2$, and $O_3$ that are singly, doubly, and triply coordinated by the Mo atoms, respectively (see Figure 1a). Therefore, three types of oxygen vacancies can be formed, which are referred to as $V_1$, $V_2$, and $V_3$. To simulate the defect models, we performed density functional theory (DFT) calculations using the Vienna *ab initio* simulation package (VASP) [41]. In the simulation, we employed the Perdew-Burke-Ernzerhof (PBE) functional [42] to optimize the crystal geometry while the Heyd-Scuseria-Ernzerhof (HSE06) hybrid functional [43] was used to obtain the electronic properties (denoted as HSE06@PBE). More details can be found in Supporting Information 15. Using the HSE06@PBE, we determined the band gap of the bulk $\alpha$-$MoO_3$ to be 3.0 eV in excellent agreement with the experimental value of 3.2 eV [44].

We calculated the formation energy of the three different vacancy sites in various charge states ($q = +2, +1, 0, -1, -2$) and found that $V_3$ is significantly unstable than the other two vacancy sites. Notably, $V_1$ and $V_2$ have very similar formation energies deviating by less than 0.2 eV (Supporting Information 15.1). The formation energy of $V_2$, as a representative, is shown in Figure 4a. Both $V_1$ and $V_2$ are found to be most stable in their +2 and +1 charge states with compensating electrons in the conduction band. This agrees well with the observed *n*-type characteristics of $\alpha$-$MoO_3$ [45]. Based on their higher stability, we further focus on $V_1$ and $V_2$ in their positive and neutral charge states for the electronic property investigations.

From our DFT calculations, we have identified the defect $V_2$ in the charge state $q = +2$, namely $V_2^{+2}$, to be the most probable origin of the observed quantum emitter (Figure 4b and details in Supporting Information 15.2). The calculated single-particle energy level of the $V_2^{+2}$ defect is shown in Figure 4c, with a singlet ground state. The energy difference between the highest occupied orbital and lowest unoccupied orbital is found to be 1.91 eV. From the calculated inverse participation ratio (IPR) that measures the degree of orbital localization, we found that the lowest unoccupied orbital is well localized while the highest occupied orbital is delocalized, or partially localized corresponding to a shallow defect state (see Figure S19) [46]. We also performed a self-consistent field calculation ($\Delta$SCF) with an electron promoted from the highest occupied to the lowest unoccupied orbital to obtain the ZPL energy and the HR factor. The calculated ZPL energy of $V_2^{+2}$ is 1.8 eV which is slightly smaller but in good agreement with experiments. The calculated HR factor of 3.59 is fully consistent with the observation of high luminosity of the optically active quantum emitters. For $V_2$ in other charge states and for the other oxygen vacancies, the calculated ZPL energies and HR factors were not compatible with the experimental observations.

Focusing on the $V_2^{+2}$ defect, we show the calculated emission spectrum [47] in Figure 4d (the calculated spectrum has been shifted to match the experimental ZPL energy). The similarity between the theoretical and experimental line shapes is striking. The most pronounced PSBs are observed within approximately 200 meV of the ZPL. The calculated electron-phonon spectral density is shown in Supporting Information 15.3. An analysis of the localized defect state (the lowest unoccupied orbital on Figure 4c) shows that it is mainly composed of d-orbitals of the Mo atoms close to the O vacancy, agreeing with a previous report [48]. We further calculated the transition dipole moment between the highest occupied state and lowest unoccupied state. The calculated dipole moment lies in the plane spanned by the *a* and *c* lattice vectors defined in Figure 1a (details in Supporting Information 15.2). We note that the observed discrepancy between experiment and theoretical results could have been affected by several factors such as local strain or edge effects which could be emitter dependent.

According to our DFT calculations, all three oxygen vacancies are singlets in their +2 charge states. However, the +1 charge states are doublets and host an unpaired electron making them interesting for applications such as magnetic field sensing or spin qubits. While the combined set of physical quantities comprised by the ZPL energy, the HR factor, and the PL line shape calculated for the +1 charge states overall show less good agreement with experiments as compared to the $V_2^{+2}$ defect (see Table S3 in the Supporting Information 15.2), we cannot fully rule out that such defects could be present in our samples. We expect that future work manipulating spin states of defects will be available through further exploration of the charge state control. Therefore, we calculated the spin coherence time ($T_2$) for a generic defect in monolayers of $\alpha$-$MoO_3$ and compared that to $MoS_2$ and *h*-BN. Our results show $T_2$ values of 2.90, 3.10, and 0.04 ms for $\alpha$-$MoO_3$, $MoS_2$, and *h*-BN, respectively. Notably, $\alpha$-$MoO_3$ exhibits a significantly long $T_2$, highlighting the potential of $\alpha$-$MoO_3$ as a host for qubit and magnetic sensing applications. The similar $T_2$ values found for $\alpha$-$MoO_3$ and $MoS_2$ are expected due to the similar nuclear spin properties of O and S.

In conclusion, we report single-photon emitters in thin layers of $\alpha$-$MoO_3$ with high emission intensity and photostability at room temperature. We found that various surface treatment methods, such as capping with polymers or graphene can be utilized to enhance the stability of the emission properties. Our DFT calculations indicate that the origin of the quantum emitters is likely to be oxygen vacancy defect sites. Moreover, spin defects in $\alpha$-$MoO_3$, if identifiable, are expected to achieve long spin coherence time due to the abundance of zero nuclear spin isotopes. With the capability of room temperature operation and possibility for electronic structure modulations, quantum emitters in $\alpha$-$MoO_3$ suits to enable various novel applications in chip-based quantum technologies. Further, through van der Waals stacking with other layered materials, $\alpha$-$MoO_3$ opens new opportunities for 2D material-based hybrid quantum nanophotonics and information science.

## Associated content

Supporting Information: Experimental methods, AFM and single-photon emitter of 5-layer $\alpha$-$MoO_3$, Single-photon emitter from $\alpha$-$MoO_3$ without annealing, Raman spectrum before and after annealing, background correction of the $g^2(\tau)$ measurement, stability and intensity test measurements at room temperature, more $g^2(\tau)$ measurements from $\alpha$-$MoO_3$, emitter with spectrometer-limited linewidth, histogram of Huang-Rhys (HR) factor, transition dynamics using a three-level model, summary of brightness of 2D single-photon emitters including $\alpha$-$MoO_3$, temperature dependent spectrum and intensity, the effect of surface treatment on the stability of defect centers, emitters from suspended $\alpha$-$MoO_3$ flake, DFT calculation of oxygen defect center (PDF)

## Acknowledgements

This work was supported by the National Research Foundation (NRF) of Korea (Grants No. 2021R1A5A103299614, No. RS-2024-00356893, No. RS-2024-00413957). Measurements were supported by the Institute for Basic Science (IBS) of Korea (Grant No. IBS-R009_D1) and the Ministry of Science and ICT (MSIT) of Korea under the Information Technology Research Center (ITRC) support program (Grant No. RS-2022-00164799). J. L. acknowledges the support from the Creative-Pioneering Researchers Program through Seoul National University. H. W. and K. S. T. acknowledge the support of Novo Nordish Foundation Challenge Programme 2021: Smart nanomaterials for applications in life-science, BIOMAG Grant NNF 21OC0066526. K. S. T. is a Villum Investigator supported by VILLUM FONDEN (Grant No. 37789).

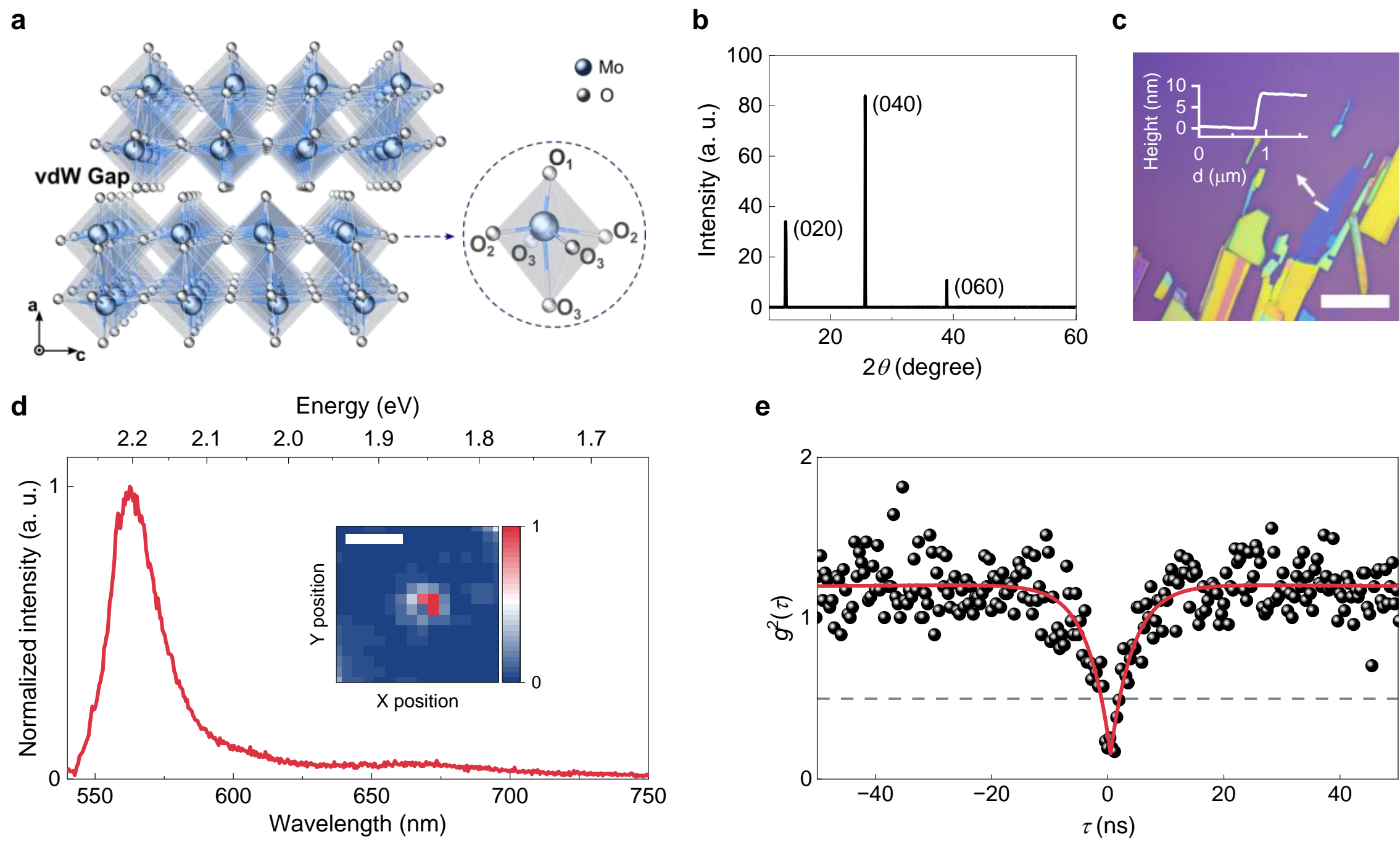

**Figure 1. $\alpha$-$MoO_3$ crystal and room temperature quantum emitter**. **(a)** Illustration of $\alpha$-$MoO_3$ lattice structure. Blue and grey balls are molybdenum and oxygen atoms, respectively. The octahedron in the right bottom panel shows the positions of three inequivalent oxygen atoms. **(b)** High resolution XRD measurement result of as-grown $\alpha$-$MoO_3$. **(c)** Optical image of an exfoliated $\alpha$-$MoO_3$ thin flake with a scale bar of 10 μm. The graph in the inset shows an AFM line profile of the flake. **(d)** Room temperature spectrum of an emitter. The inset shows the spatially localized emission profile with a scale bar of 3 μm. **(e)** Second-order correlation function of the emitter shown in **(d)**. A red solid line represents the fitting result. A dashed horizontal line at 0.5 is shown to provide the guide to the criterion for single-photon emission. During the measurement, the laser power was maintained at 5 μW.

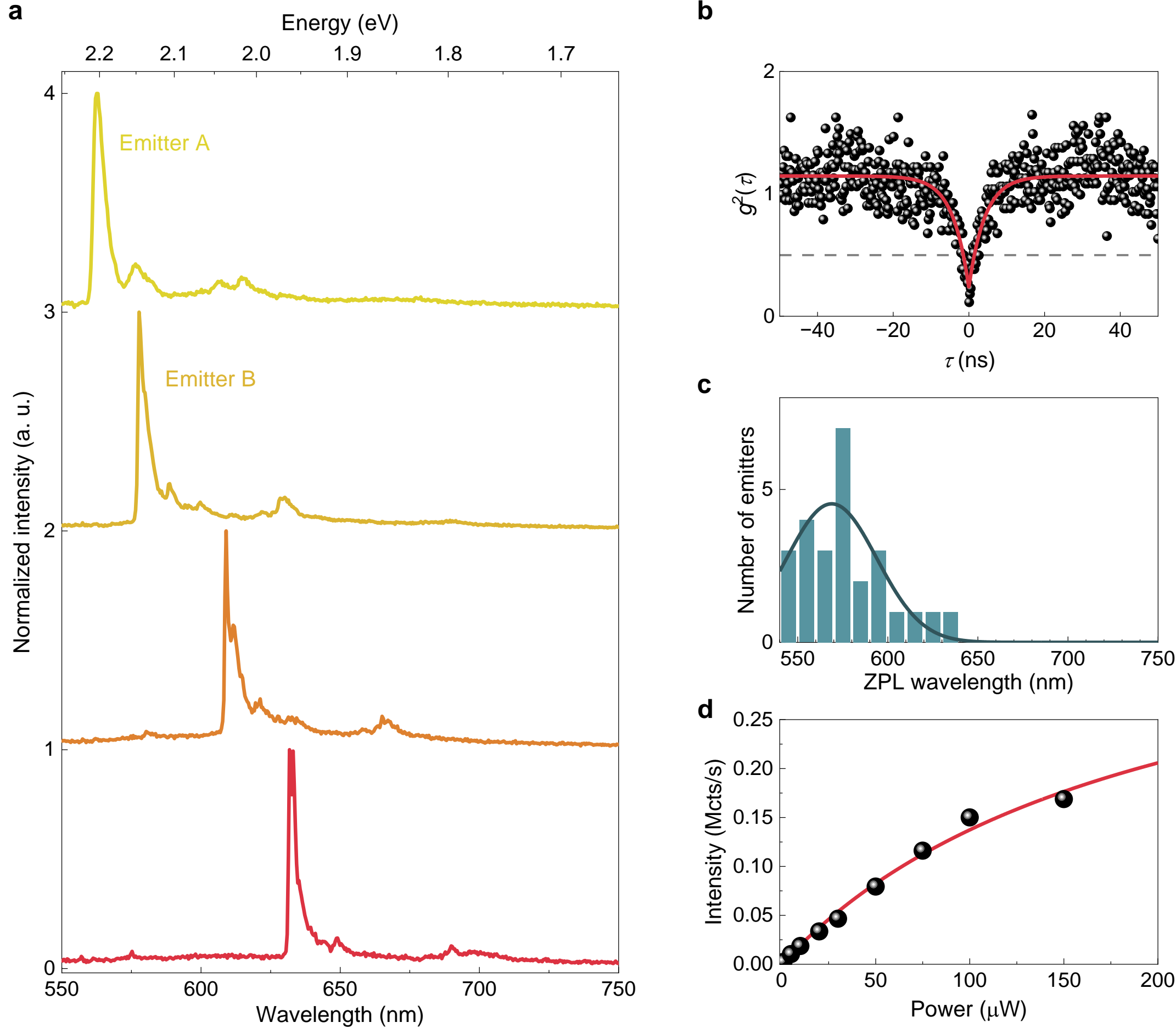

**Figure 2. Emission properties of quantum emitters in $\alpha$-$MoO_3$ at 10 K. (a)** Photoluminescence spectrum of several emitters in $\alpha$-$MoO_3$ measured at 10 K. The spectrum was acquired using a laser excitation power of 10 μW. **(b)** Second-order correlation function of emitter B in **(a)**. A red solid line represents the fitting result. **(c)** ZPL energy distribution of emitters. A Gaussian fit is shown as a solid line. **(d)** Emission intensity of emitter A as a function of laser power with the fitting result depicted as a red solid line.

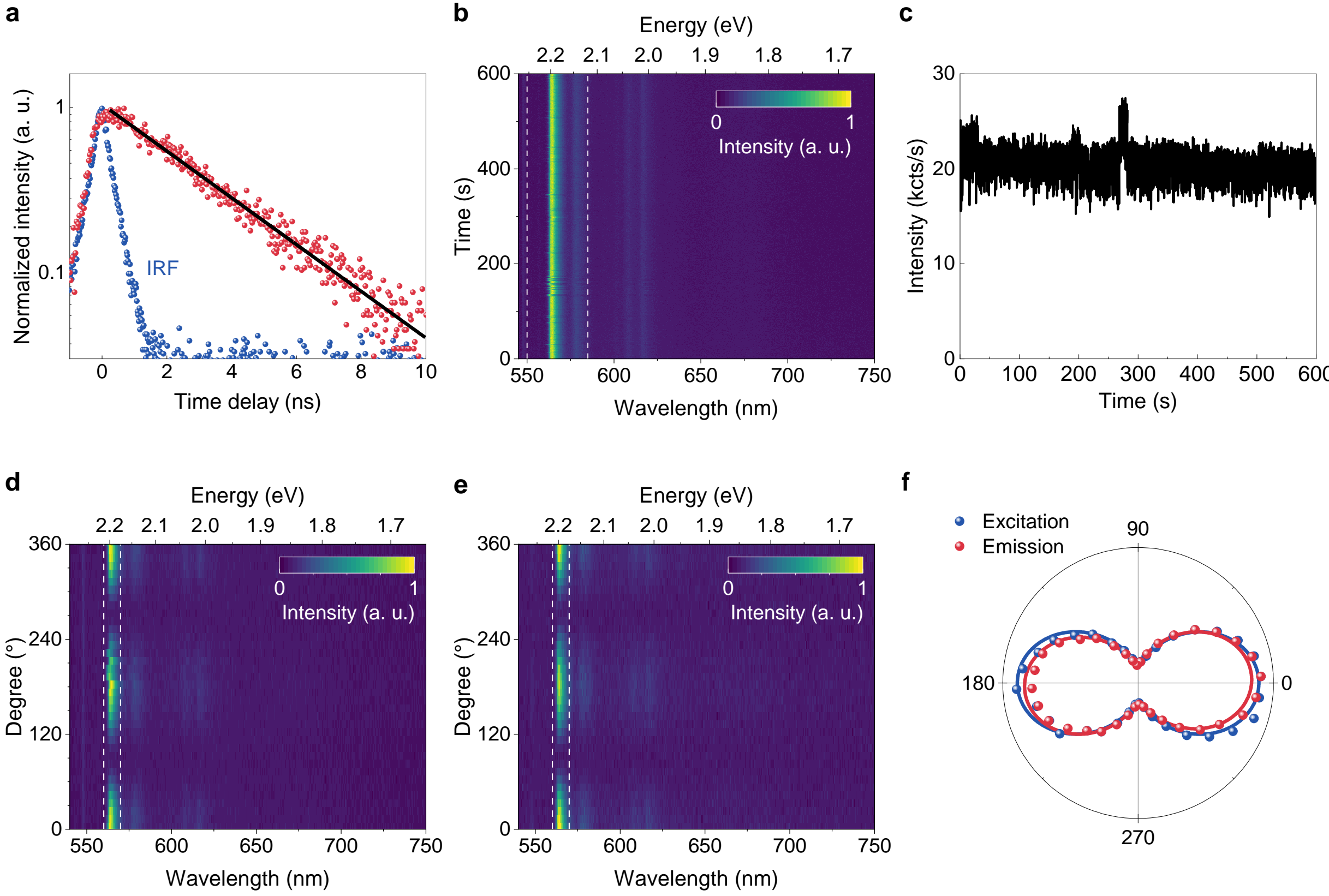

**Figure 3. Emission stability and polarization dependence of quantum emitters. (a)** Time-resolved lifetime measurement result of an emitter (red dot) measured at 10 K. The power of the pulse laser was 2.5 μW. The exponential fit result (black solid line) provides an excited state lifetime of 3.07 ns. Blue dot represents the instrument response function (IRF). **(b)** Stability measurement of the emission intensity as a function of both time and wavelength over a duration of 10 mins. **(c)** Fluorescence stability monitored with SPCM after filtering the wavelength range indicated by the white dashed lines in **(b)**. **(d, e)** Polarization dependence measurement of ZPL and phonon side bands as a function of excitation laser polarization **(d)** and emission polarization **(e)**, respectively. **(f)** Polar mapping of ZPL intensity for excitation laser polarization (blue dot) and emission polarization (red dot) dependence. Solid lines are $I(\theta) = A\sin^2(\theta - \theta_0) + B$ fit result to each data.

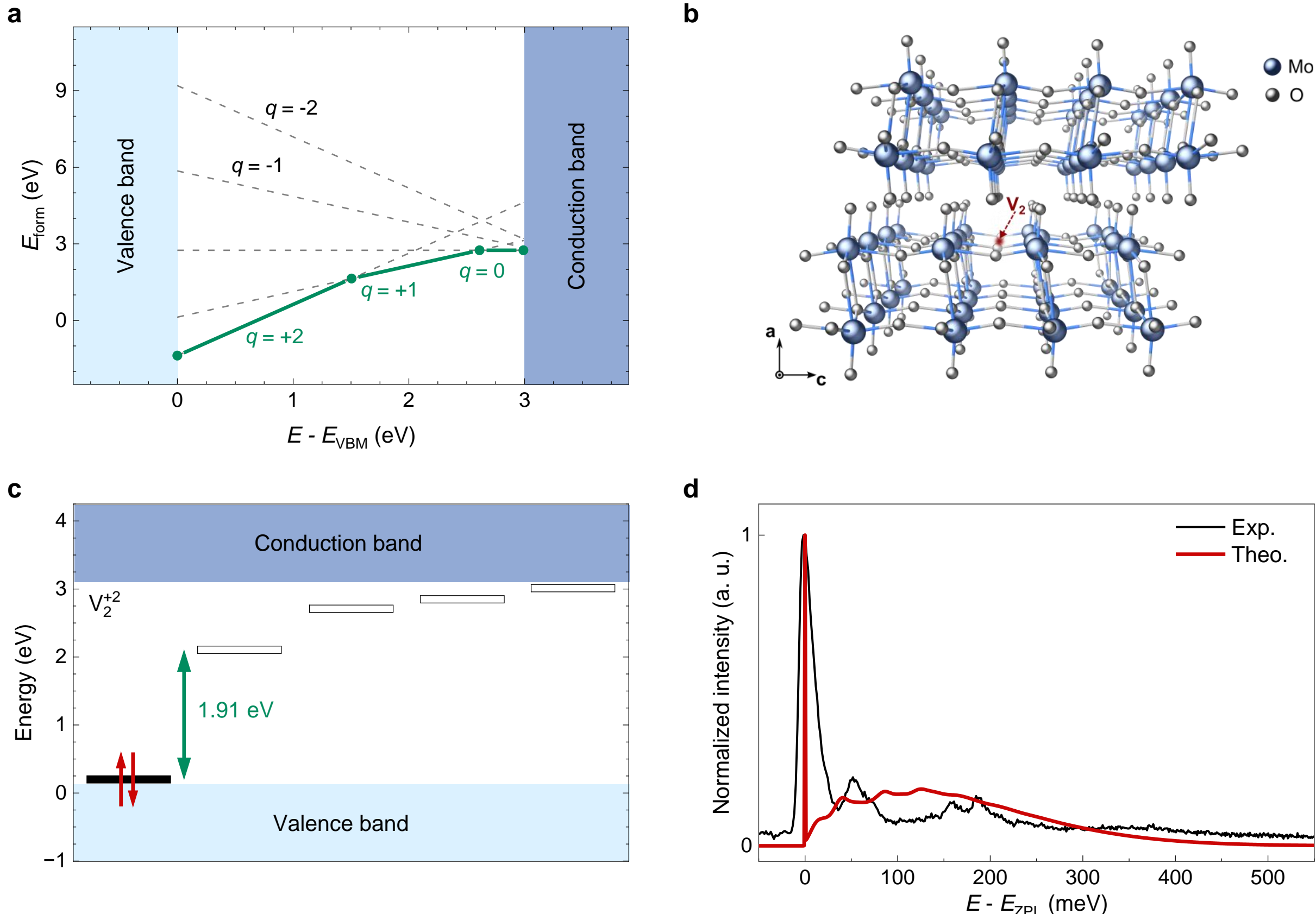

**Figure 4. DFT calculation of oxygen vacancy defect center. (a)** Formation energy of $V_2$ defect as a function of the Fermi energy ($E_F$). Each dashed line represents charge states encompassing +2 to -2. The green solid line highlights the most stable charge state along the $E_F$. **(b)** The crystal structure of $V_2^{+2}$ defect. **(c)** Kohn-Sham energy levels of $V_2^{+2}$ defect. The filled and empty blocks stand for the occupied and unoccupied states, respectively, with red arrows indicating electron spins. The energy difference between the highest occupied and lowest unoccupied state is 1.91 eV. **(d)** Theoretical calculation of $V_2^{+2}$ emission spectrum (red solid line). Experimentally observed spectrum (emitter A) is also shown (black solid line). The energy of both spectra is shifted horizontally from the ZPL energy to compare the shape and distribution of phonon side bands.

## TOC graphic

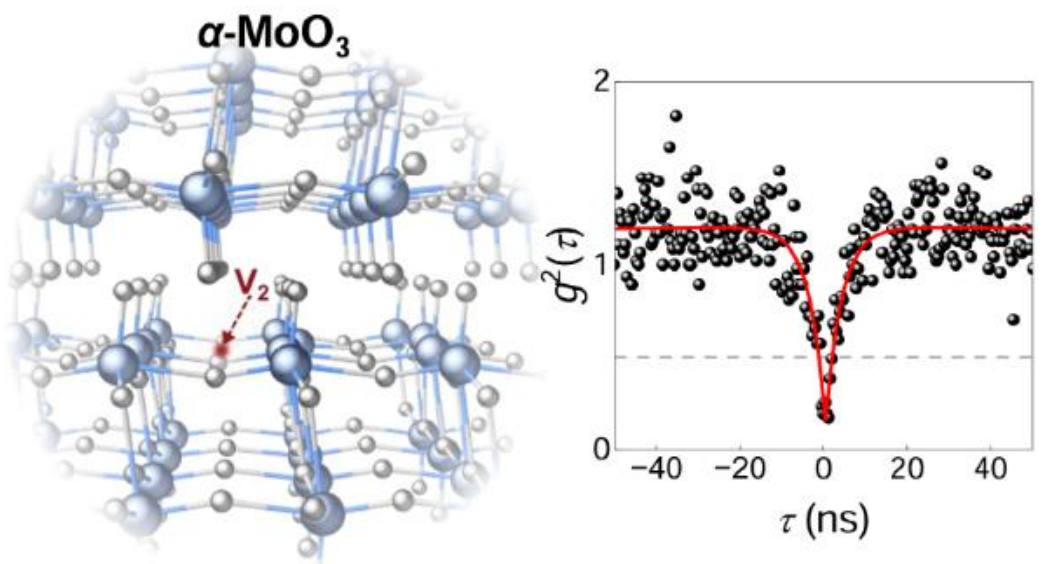